\documentclass[preprint,preprintnumbers,amsmath,amssymb]{revtex4}
\usepackage{bbm}
\usepackage{graphicx}
\usepackage{color}

\newcommand{\norm}[1]{\left\Vert#1\right\Vert}

\newcommand{\abs}[1]{\left\vert#1\right\vert}

\newcommand{\ket}[1]{\vert #1 \rangle}
\newcommand{\amp}[2]{\left<#1 \vert #2 \right>}

\begin{document}
\title{Continuous-Time Quantum Random Walks Require Discrete Space}
\author{K. Manouchehri}
\email{kia@physics.uwa.edu.au}
\author{J.B. Wang}
\email{wang@physics.uwa.edu.au} \affiliation{School of Physics, The
University of Western Australia}
\date{\today}

\begin{abstract}
    Quantum random walks are shown to have non-intuitive
    dynamics which makes them an attractive area of study for
    devising quantum algorithms for long-standing open problems as well
    as those arising in the field of quantum computing. In the case
    of continuous-time quantum random walks, such peculiar
    dynamics can arise from simple evolution operators closely resembling
    the quantum free-wave propagator. We investigate the divergence of
    quantum walk dynamics from the free-wave evolution
    and show that in order for continuous-time quantum walks to display their
    characteristic propagation, the state space must be
    discrete. This behavior rules out many continuous quantum
    systems as possible candidates for implementing
    continuous-time quantum random walks.
\end{abstract}


\maketitle

\section{Introduction}

Quantum random walks represent a generalized version of the well
known classical random walk, which can be elegantly described using
quantum information processing terminology \cite{Aharonov1993}.
Despite their apparent connection however, dynamics of quantum
random walks are often non-intuitive and deviate significantly from
those of their classical counterparts \cite{Farhi1998}. Among the
differences, the faster mixing and hitting times of quantum random
walks are particularly noteworthy, making them an attractive area of
study for devising efficient quantum algorithms, including those
pertaining to connectivity and graph theory \cite{Kempe2003,
Farhi1998, Childs2003}, as well as quantum search algorithms
\cite{Shenvi2003, Childs2004}.

There are two broad classes of quantum random walks, namely the
discrete- and continuous-time quantum random walks, which have
independently emerged out of the study of unrelated physical
problems. Despite their fundamentally different quantum dynamics
however, both families of walks share similar and characteristic
propagation behavior \cite{Kempe2003, Konno2005, Patel2005}.
Strauch's recent work \cite{Strauch2006} is the latest in a line of
theoretical efforts to establishing a formal connection between the
discrete and continuous-time quantum random walks, in a manner
similar to their classical counterparts.

In this paper we investigate the discretization of space in
continuous-time quantum walks. To the best of our knowledge is the
first study examining the relationship between discrete- and
continuous- \emph{space} quantum random walks. In what follows we
present, in Sec. \ref{section.cont-qrw-overview}, an introductory
overview of continuous-time quantum random walks. Then in Sec.
\ref{section.cont-vs-disc} we provide a concise definition of
discrete and continuous state space, and in Sec
\ref{section.cont-disc-trans-model} describe the way in which we
model the transition of the state space from discrete to continuous.
We present our results in Sec. \ref{section.results} and show how
such a transition can reduce the quantum walk evolution to the
typical quantum wave propagation in free space. In conclusion we
discuss the implications of our findings for the physical
implementation of continuous-time quantum random walks. In
particular we argue that existing experimental implementations
\cite{Du2003, Cote2006, Solenov2006} implicitly verify our assertion
that continuous-time quantum random walks require discrete space.

\section{Continuous-time Quantum Random Walk Overview}
\label{section.cont-qrw-overview}

Continuous-time quantum random walks were initially proposed by
Farhi and Gutmann \cite{Farhi1998} in 1998, out of a study of
computational problems formulated in terms of decision trees.
Suppose we are given a decision tree that has $N$ nodes indexed by
integers $i=1,\ldots,N$. We can then define an $N\times N$
transition rate matrix $\mathcal{H}$ with elements
\begin{equation}
    \begin{array}{c}
        h_{ij} = \left\{
        \begin{array}{ll}
            -\gamma_{ij} & \text{for $i \neq j$ if node $i$ is connected to node $j$}  \\
            0 & \text{for $i \neq j$ if node $i$ is not connected to node $j$} \\
            S_i  & \text{for $i = j$}
        \end{array} \right.
        \label{eqn.trans-rule-def}
    \end{array},
\end{equation}
where $\gamma_{ij}$ is the probability per unit time for making a
transition from node $i$ to node $j$ and for $\mathcal{H}$ to be
conservative
\begin{equation}\label{eqn.trans-mat-conservative}
    S_i = \sum_{\substack{j=1\\j \neq i}}^N \gamma_{ij}.
\end{equation}
Defining $\mathbf{P}$ as the probability distribution vector for the
nodes, the transitions can be described by
\begin{equation}
    \frac{d \mathbf{P}(t)}{d t} = -\mathcal{H} \mathbf{P}(t),
\end{equation}
for which the solution is
\begin{equation}\label{eqn.qrw-cont-eqn}
    \mathbf{P}(t) = \exp(-\mathcal{H} t) \mathbf{P}(0),
\end{equation}
known as the master equation.

Farhi and Gutmann's contribution was to propose using the
classically constructed transition rate matrix $\mathcal{H}$ to
evolve the continuous-time state transitions \emph{quantum
mechanically}. This involved replacing the real valued probability
distribution vector $\mathbf{P}(t)$ with a complex valued wave
vector $\psi(t)$ and adding the complex notation $\mathbbmtt{i}$ to
the evolution exponent, i.e.
\begin{equation}\label{eqn.qrw-evolution}
    \psi(t) = \exp(-\mathbbmtt{i} \mathcal{H} t) \psi(0).
\end{equation}
Hence the probability distribution vector
$\mathbf{P}(t)=\abs{\psi(t)}^2$ and the elements of $\psi(t)$ are
the complex amplitudes $\psi(i,t)=\amp{i}{\psi(t)}$ where
$\ket{\psi(t)}$ is the state of the entire decision tree system at
time $t$. In this quantum evolution the transition matrix
$\mathcal{H}$ is required to be Hermitian. This formulation is not
limited to discission trees and can be readily applied to the
continuous-time quantum random walk on any general undirected graph
with $N$ vertices.

\begin{figure}[h]
    \centering
    \includegraphics[width=10cm]{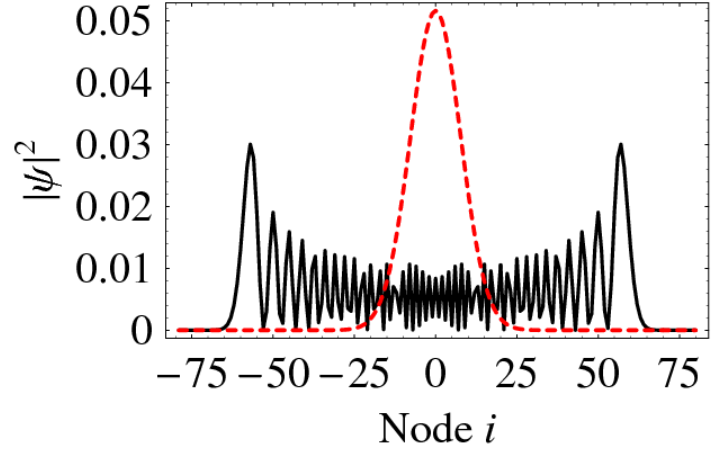}
    \caption{Continuous-time quantum random walk probability distribution (solid) using the transition rate matrix in Eq. \ref{eqn.trans-rate-matrix}, with initial condition $\psi(i=0,t=0)=1$ and evolution time $t=15$. Also plotted (dashed) is the time evolution of a continuous-time \emph{classical} random walk using the same transition rate matrix, initial condition $\mathbf{P}(i=0,t=0)=1$ and evolution time $t=25$.}
    \label{fig.qrw-vs-crw}
\end{figure}

Figure \ref{fig.qrw-vs-crw} shows the characteristic probability
distribution $\abs{\psi(t=15)}^2$ of a continuous-time quantum
random walks on a line with $N=160$ nodes, indexed by $i = -79
\ldots 80$, and the initial state $\psi(i=0,t=0)=1$. For this
quantum walk each node is assumed to be connected only to its
neighboring nodes by a constant transition rate $\gamma=1$ resulting
in a transition rate matrix given by
\begin{equation}
    \mathcal{H} = \left(
    \begin{array}{ccccc}
        -2 & 1 & 0 & 0 &  \\
        1 & -2 & 1 & 0 &  \\
        0 & 1 & -2 & 1 & \cdots \\
        0 & 0 & 1 & -2 &  \\
        & & \vdots & & \ddots
    \end{array}\right).
    \label{eqn.trans-rate-matrix}
\end{equation}
For comparison we have also plotted the continuous-time
\emph{classical} random walk probability distribution
$\mathbf{P}(t=15)$ using the same transition rate matrix and initial
condition $\mathbf{P}(i=0,t=0)=1$.

\section{Discrete vs Continuous State Space}
\label{section.cont-vs-disc}

\begin{figure}
    \centering
    \includegraphics[width=10cm]{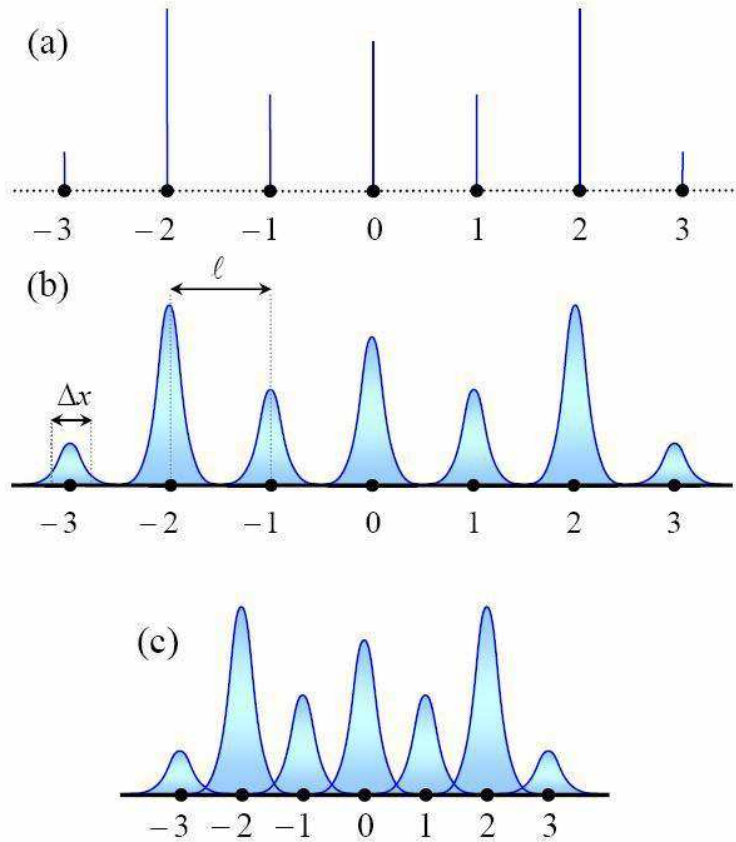}
    \caption{a) Continuous-time quantum random walk on a line of discrete nodes. Vertical lines represent the walker's probability amplitude to be at each node. d) Continuous-time quantum random walk on a continuous line. The nodes of the walk are a distance $\ell$ apart and localized distributions represent the walker's probability amplitude to be at each node with an associated uncertainty $\Delta x$. The condition $\ell \gg \Delta x$ means that the nodes are well separated and their respective distributions do not interfere. c) The nodes of the walk have been moved closer such that $\ell \sim \Delta x$. Consequently the localized distributions at neighboring nodes begin to overlap and interfere with one another.}
    \label{fig.qrw-state-space}
\end{figure}

In Farhi and Gutmann's treatment of the quantum walk, an arbitrary
graph with $N$ vertices can be represented as $N$ \emph{position
states} with coordinate vectors $\ket{\vec{x}_i}$, for $i=1,2\ldots
N$. These state vectors form an orthonormal basis in the
$N$-dimensional Hilbert space $\mathcal{S}$, that is
$\amp{\vec{x}_i}{\vec{x}_j}=\delta_{ij}$ and the wavefunction
remains normalized. The time evolution of the quantum walk can be
considered as continuously displacing the walker (in time) by a
distance $\ell_{ij}$ from node $i$ to all its neighboring nodes $j$
at the rate $\gamma_{ij}$, where $\ell_{ij}=\norm{\vec{x}_i -
\vec{x}_j}$ is defined as the transition length. Since the quantum
walker can only be present at positions $\vec{x}_1, \vec{x}_2 \ldots
\vec{x}_N$, we say that the state space is discrete and the walker
has an infinitely narrow width. In other words there is no
uncertainty or distribution associated with the amplitude
\begin{equation}
    \psi(i,t)=\amp{\vec{x}_i}{\psi(t)}.
\end{equation}
A simple example of this is the quantum walk on a line, illustrated
in Fig. \ref{fig.qrw-state-space}a, where the position states are
discrete nodes arranged in a line and the amplitudes $\psi(i,t)$ are
diagrammatically represented as narrow lines over each node. Clearly
in this discrete model the nodes can be made arbitrarily close by
making $\ell \longrightarrow 0$ without affecting the outcome in any
way.

The situation changes however when the state space is continuous,
meaning that vector $\vec{x}$ is no longer restricted to coordinates
$\vec{x}_1, \vec{x}_2 \ldots \vec{x}_N$ and the Hilbert space is
an infinite dimensional continuum $\mathcal{S}'$. A
consequence of this is that the quantum walker's position at
$\vec{x}_i$ can now have a finite uncertainty associated with it.
This is conveniently represented as a distribution with a finite width
$\Delta x_i$ that is centered at $\vec{x}_i$. Taking our previous
example of a quantum walk on a line, this continuous model is
depicted in Fig. \ref{fig.qrw-state-space}b where the
amplitude $\psi(i,t)$ is given by the area under the Gaussian
distribution centered at $\vec{x}_i$.

Despite the continuous nature of the state vectors in
$\mathcal{S}'$, we can nevertheless use this system to simulate the
quantum walk in the discrete Hilbert space $\mathcal{S}$. To
implement this \emph{virtually-discrete state space} over a
continuous one, we simply ensure that the states $\ket{\vec{x}_1},
\ket{\vec{x}_2} \ldots \ket{\vec{x}_N}$ remain orthogonal by
requiring that the transition length $\ell$ between all the nodes or
position states of the walk obeys $\ell \gg \Delta x$. In doing so
the overlap between the distributions at the neighboring nodes
becomes negligible, which is the case in Fig.
\ref{fig.qrw-state-space}b.

What we propose here is that the relationship $\ell \gg \Delta x$ is
in fact a necessary condition for the quantum random walk to retain its
characteristic features. In other words quantum random walks require a
discrete or orthonormal state space. In particular we show that for
a continuous-time continuous-space quantum walk on a line, where the
walker has a finite width, the evolution of the walk is
reduced to the conventional quantum wave propagation, in the limit where the transition length
approaches a continuum, i.e. $\ell \longrightarrow 0$ (see Fig.
\ref{fig.qrw-state-space}c).

In what follows we use the finite difference approximation to
construct an arbitrary transition rate matrix $\mathcal{H} \approx
-\frac{1}{2} \nabla^2$ for the quantum walk on a line. We then show
that for $\ell \gg \Delta x$ the time evolution results in the
characteristic quantum random walk signature as expected. But as the
transition length $\ell$ is reduced while keeping $\Delta x$
unchanged, the propagation behavior begins to alter until it
converges to that of a quantum wave in free space.

\section{Modeling the Discrete to Continuous Transition}
\label{section.cont-disc-trans-model}

\begin{figure}
    \centering
    \includegraphics[width=10cm]{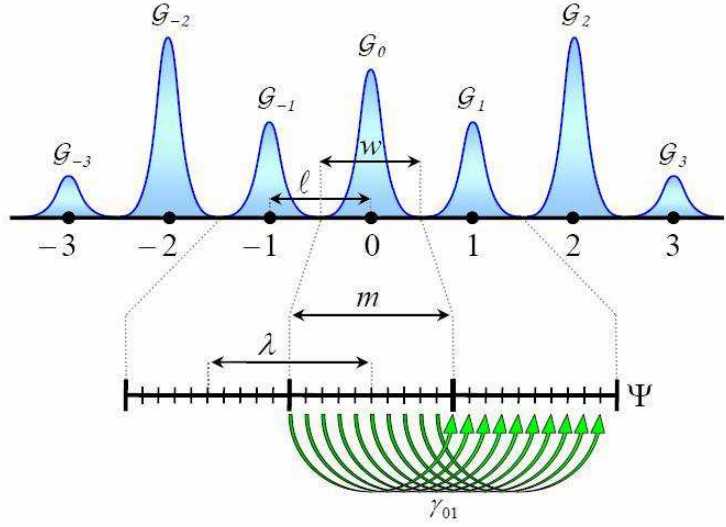}
    \caption{A virtually-discrete quantum walk on a continuous line. The line is divided into segments of length $w$ containing the localized distributions and the transition length between neighboring nodes is $\ell$. This continuous state space is then represented on a numerical vector $\Psi$ with integers $m$ and $\lambda$ corresponding to parameters $w$ and $\ell$ respectively. A transition from node $i$ to node $j$ involves making a transition from all the elements within the $i$th segment to the $j$th segment of the vector.}
    \label{fig.qrw-virtually-discrete}
\end{figure}

We start with the continuous state space in Fig.
\ref{fig.qrw-virtually-discrete} where the continuous position state
vector $\vec{x}$ is formally equivalent to a variable $x \in
\mathbb{R}$ along a line. This continuous position space is used to
construct a virtually-discrete state space quantum walk on a line by
ensuring that the condition $\ell \gg \Delta x$ is satisfied. The
line is then broken up to adjacent segments of width $w$, with the
position of the equidistant nodes $x_1, x_2 \ldots x_N$ at the
center of each segment. The natural uncertainty in the quantum
walker's position at $x_i$ is given by a Gaussian distribution
\begin{equation}\label{eqn.distribution-amplitude}
    \mathcal{G}_i(x,t) = \mathcal{A}_i(t)
    \exp\left((x-x_i)^2/(2 \Delta x)^2\right),
\end{equation}
where $\mathcal{A}_i(t)$ is a complex phase. The amplitude of the
walker to be at position $x_i$ is then given by
\begin{equation}\label{eqn.distribution-amplitude-sum}
    \psi(i,t) = \int_{L_{\min}}^{L_{\max}} \mathcal{G}_i(x,t) ~ dx
\end{equation}
where $L_{\max}=x_i + w/2$ and $L_{\min}=x_i - w/2$ are the upper
and lower bounds of the block containing the distribution. The
condition $w \gg \Delta x$, guarantees that the distribution only
finds appreciable values inside this block and is approximately zero
elsewhere.

\begin{figure}
    \centering
    \includegraphics[width=8cm]{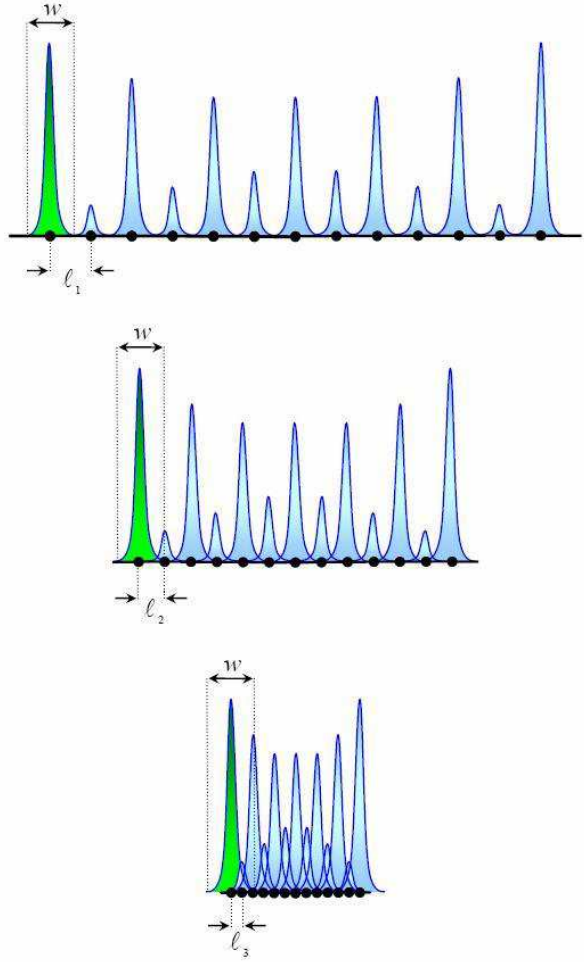}
    \caption{Moving the nodes closer (i.e. $\ell_1 \rightarrow \ell_2 \rightarrow \ell_3 \ldots$) while leaving $w$ unchanged emulates a move from a virtually-discrete to continuous space. This results in the overlap and interferant of the Gaussian distributions at the neighboring nodes.}
    \label{fig.qrw-disc-to-cont}
\end{figure}

In the way we have defined the adjacent segments, the
width $w$ is equivalent to the transition length $\ell$. What we
want to model now is the effect of reducing $\ell$ without changing
$w$, which causes the neighboring distributions to overlap as the
nodes come closer (see Fig. \ref{fig.qrw-disc-to-cont}).
Keeping $w$ constant is justified as the distribution within the
segments is associated with some fundamental uncertainty in the
position state which is unaffected by the change in the transition
length. The overlap region obviously grows as the nodes get closer
while the state space itself (i.e. the line) shrinks as a result. As
$\ell \longrightarrow 0$ and the virtually-discrete state space
approaches a continuum, it is no longer orthogonal due to the
overlapping segments, i.e. $\abs{\amp{x_i}{x_j}} > 0$ and
consequently amplitudes $\psi(i,t)$ as given by Eq.
\ref{eqn.distribution-amplitude-sum} involve summations over some
mutual areas.

In order to numerically evolve the quantum walk, we represent the
continuous state space of the walk on a finite dimensional complex
vector $\Psi$ of length $N \times \lambda$, where $\lambda$ is the
integer equivalent of $\ell$, representing the number of elements
between $\kappa_i$ and $\kappa_{i+1}$, and $\kappa_1, \kappa_2
\ldots \kappa_N$ are elements of the vector corresponding to the
node positions $x_1, x_2 \ldots x_N$ along the continuous line. We
also introduce $m$ as the integer equivalent of $w$, denoting the
number of vector elements across each segment.

\begin{figure}
    \centering
    \includegraphics[width=8cm]{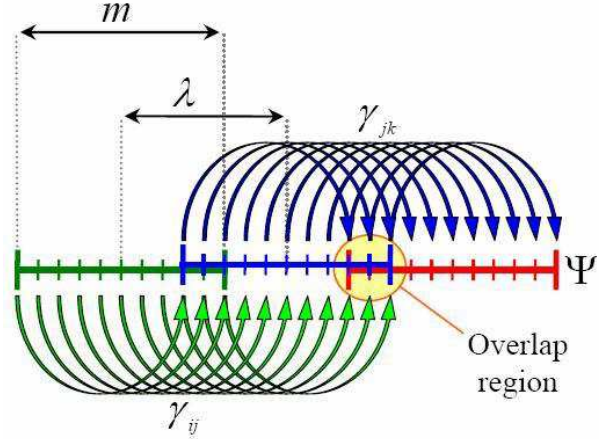}
    \caption{Approaching the nodes of the quantum walk represented on vector $\Psi$ involves reducing $\lambda$ while leaving $m$ unchanged which leads to an overlap region where vector elements are shared by neighboring distributions. A transition from node $i$ to node $j$ involves making a transition from all the elements within the $i$th segment to the $j$th segment of the vector. Vector elements in the overlap region are subject to the cumulative transition of the overlapping nodes.}
    \label{fig.qrw-overlapping-segments}
\end{figure}

Given an arbitrary $N \times N$ transition rate matrix
$\mathcal{H}$, in order to propagate the quantum walk in the
quasi-continuous space represented by vector $\Psi(t)$, we introduce
a modified matrix $\mathbf{H}$ of size $N \lambda \times N \lambda$.
To construct this matrix, we note that a transition from the $i$th
to the $j$th node in the discrete walk corresponds to the transition
of all the elements within segment $i$ to segment $j$ as depicted in
Fig. \ref{fig.qrw-virtually-discrete} and
\ref{fig.qrw-overlapping-segments}. This can be represented by a
\emph{block} matrix
\begin{equation}
    \Gamma_{ij} = \left(
    \begin{array}{ccc}
        0 & 0 & 0  \\
        0 &\mathbf{B}_{ij} & 0 \\
        0 & 0 & 0  \\
    \end{array}\right)_{N \lambda \times N \lambda},
\end{equation}
which is zero everywhere except in an $m \times m$ block
\begin{equation}
    \mathbf{B}_{ij} = \gamma_{ij} ~ \mathcal{I}_{m \times m},
\end{equation}
which is centered at $\kappa_i$ and $\kappa_j$, and $\mathcal{I}$
represents the identity matrix. The modified transition matrix is
then given by
\begin{equation}
    \mathbf{H} = \sum_{i=1}^N \sum_{j=1}^N \Gamma_{ij},
\end{equation}
and the continuous-time quantum random walk proceeds according to
\begin{equation}\label{eqn.ext-qrw-cont-eqn}
    \Psi(t) = \exp(-\mathbbmtt{i} \hat{\mathbf{H}} t) \Psi(0).
\end{equation}

\section{Results}
\label{section.results}

In our simulations we considered two quantum walks, both with $N =
160$ discrete nodes (enumerated as -79, ... 0, ... 80) arranged on a
line, and transition matrices $\mathcal{H}_1$ and $\mathcal{H}_2$
which were constructed as the finite difference approximations to
$-\frac{1}{2}\nabla^2$. For $\mathcal{H}_1$ we use the 1$^\text{st}$
order approximation, where
\begin{equation} \label{eqn.trans-rule-def-1}
    \nabla^2 \psi(i) \approx \frac{1}{2 h^2}(\psi(i-1) - 2\psi(i) +
    \psi(i+1)).
\end{equation}
Here we set grid spacing $h = 1$, and the time parameter in $\psi$
is implicit. Transition rates $\gamma_{ij}$ are the coefficients of
$\psi(i-s)$ divided by -2, where $i - j = s$ and zero
otherwise. $\mathcal{H}_2$ is constructed similarly but using the
10$^\text{th}$ order approximation, where
\begin{eqnarray} \label{eqn.trans-rule-def-2}
    \nabla^2 \psi(i) \approx \frac{1}{25200 h^2} && (8\psi(i-5) -125\psi(i-4) +\\
    \nonumber && ~1000\psi(i-3) -6000\psi(i-2) +\\
    \nonumber && ~42000\psi(i-1) -73766\psi(i) + 42000\psi(i+1) + \\
    \nonumber && ~1000\psi(i+3) - 6000\psi(i+2) + \\
    \nonumber && ~8\psi(i+5) - 125\psi(i+4)).
\end{eqnarray}
For each walk we then defined the vector $\Psi$ using $m = 16$ which
we initialized so that $\Psi(t=0)$ is zero everywhere except for a
single distribution $\mathcal{G}_{i=0}(x,t=0)$ (Eq.
\ref{eqn.distribution-amplitude}), corresponding to $\psi(i=0,
t=0)=1$. The state space transition from discrete to continuous was
then simulated by computing the quantum walk evolution for 5
different values of $\lambda = 16, 4, 3, 2$ and 1.

In simulating the walk it is possible to construct the modified
transition matrices $\mathbf{H}_1$ and $\mathbf{H}_2$ as described
in the previous section and compute the time evolution from Eq.
\ref{eqn.ext-qrw-cont-eqn}. The evaluation of matrix exponentials
however is computationally expensive and even for modest choices of
$N$ and $m$, large matrices have to be stored, and the evaluation
time becomes prohibitively long, rendering this \emph{direct method}
impractical. Instead in the Appendix we have introduce an
alternative computational scheme based on Fourier analysis, which
provided us with a dramatic improvement in the computation time.

\begin{figure}
    \centering
    \includegraphics[width=5cm]{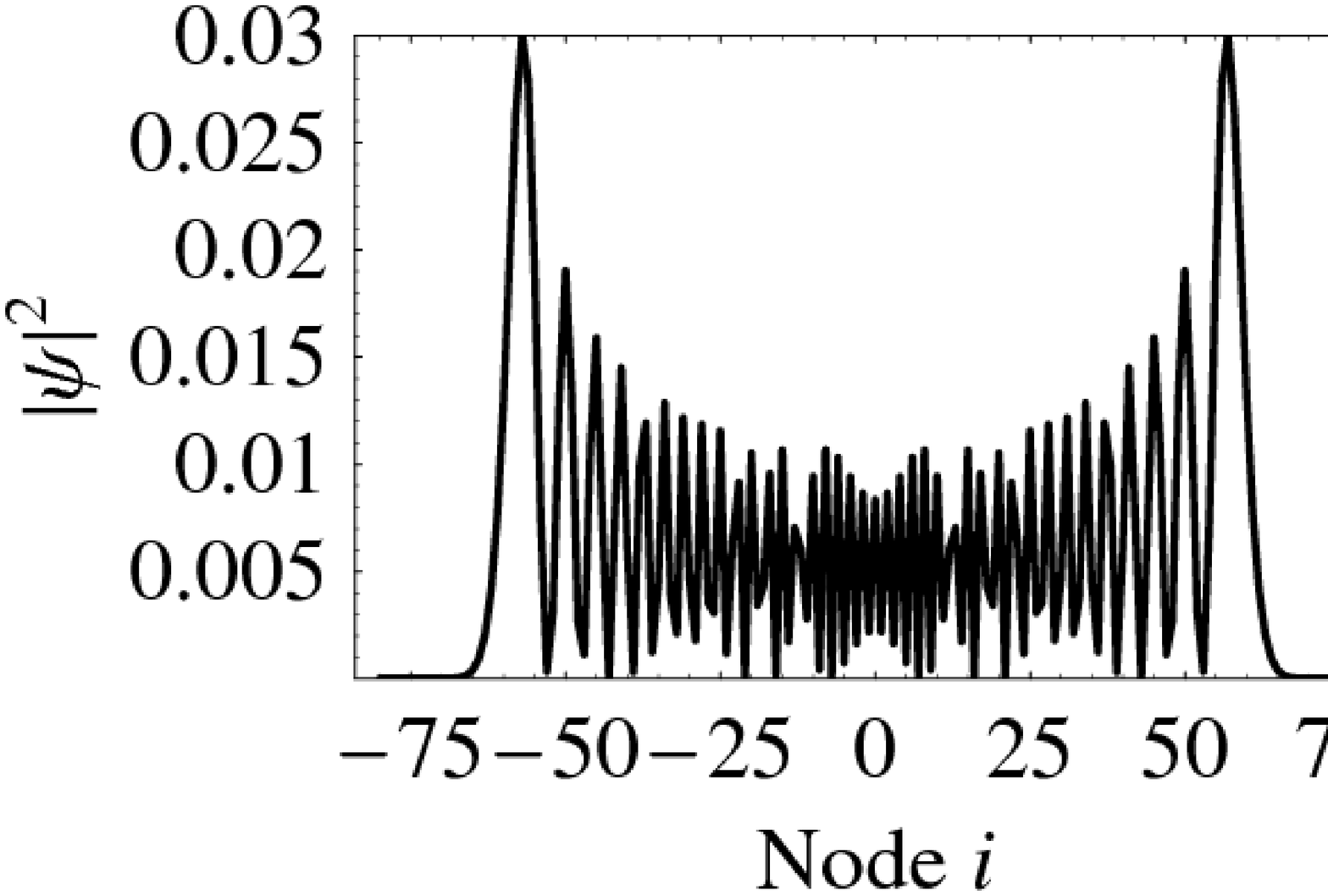}
    \includegraphics[width=5cm]{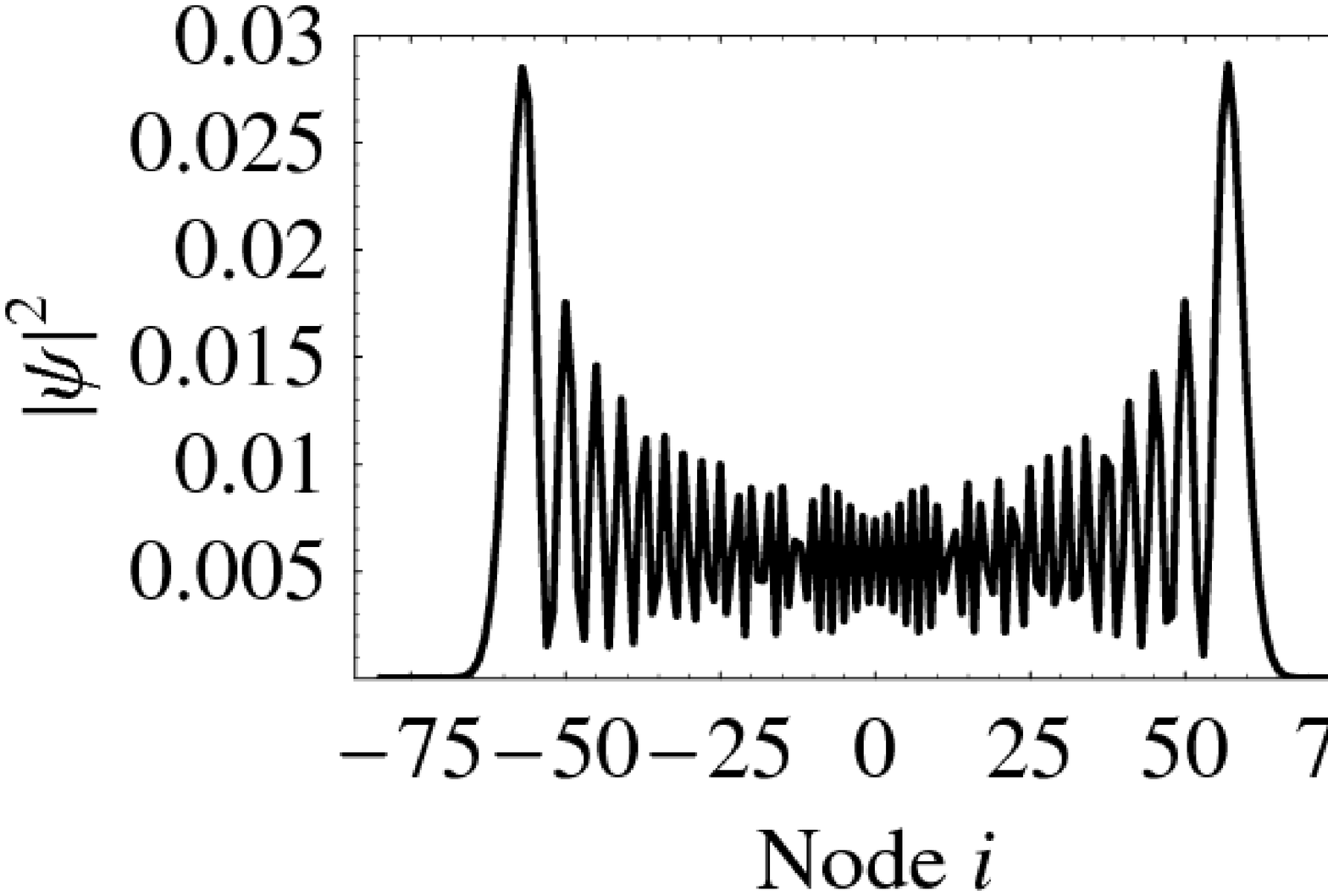}
    \includegraphics[width=5cm]{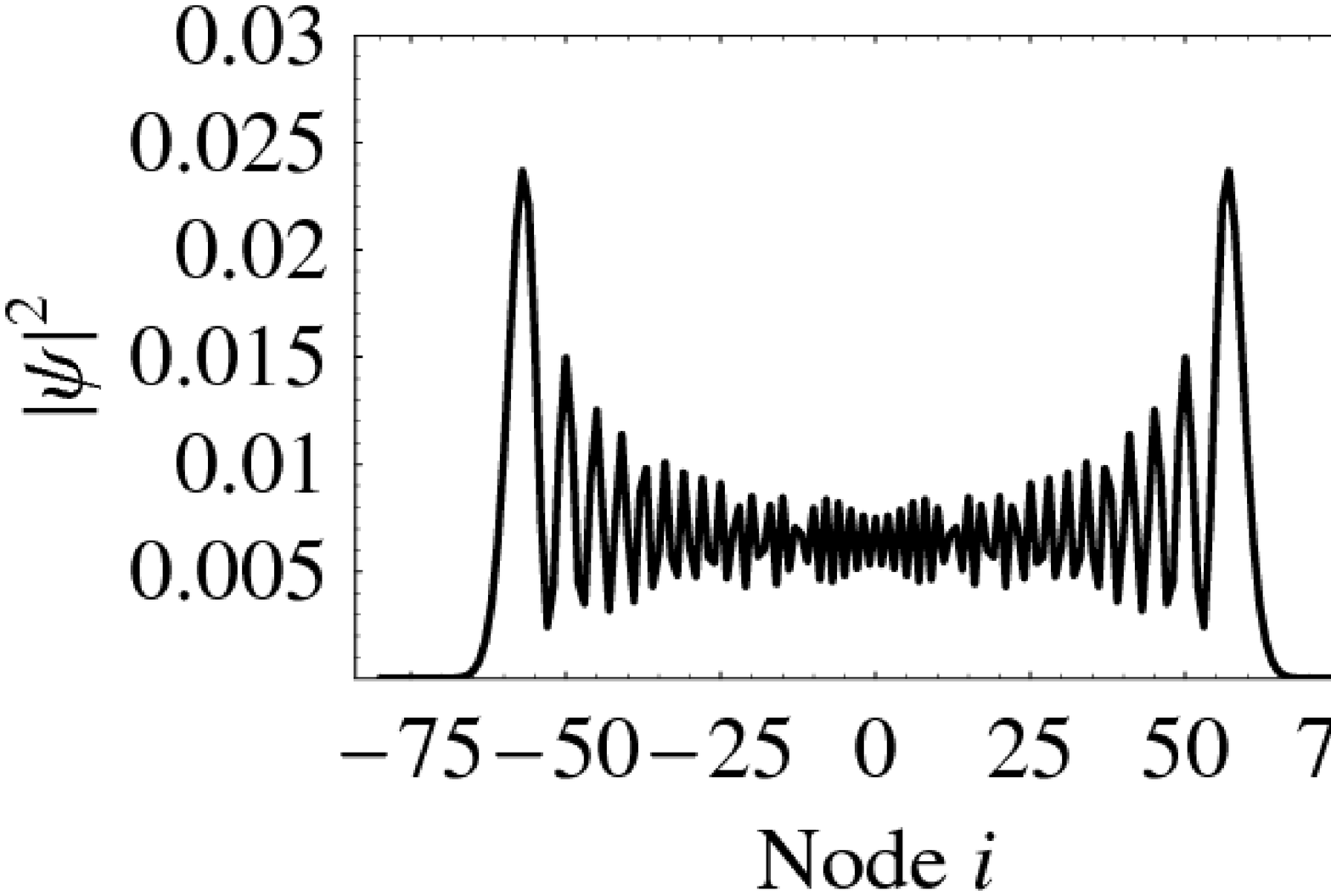}
    \includegraphics[width=5cm]{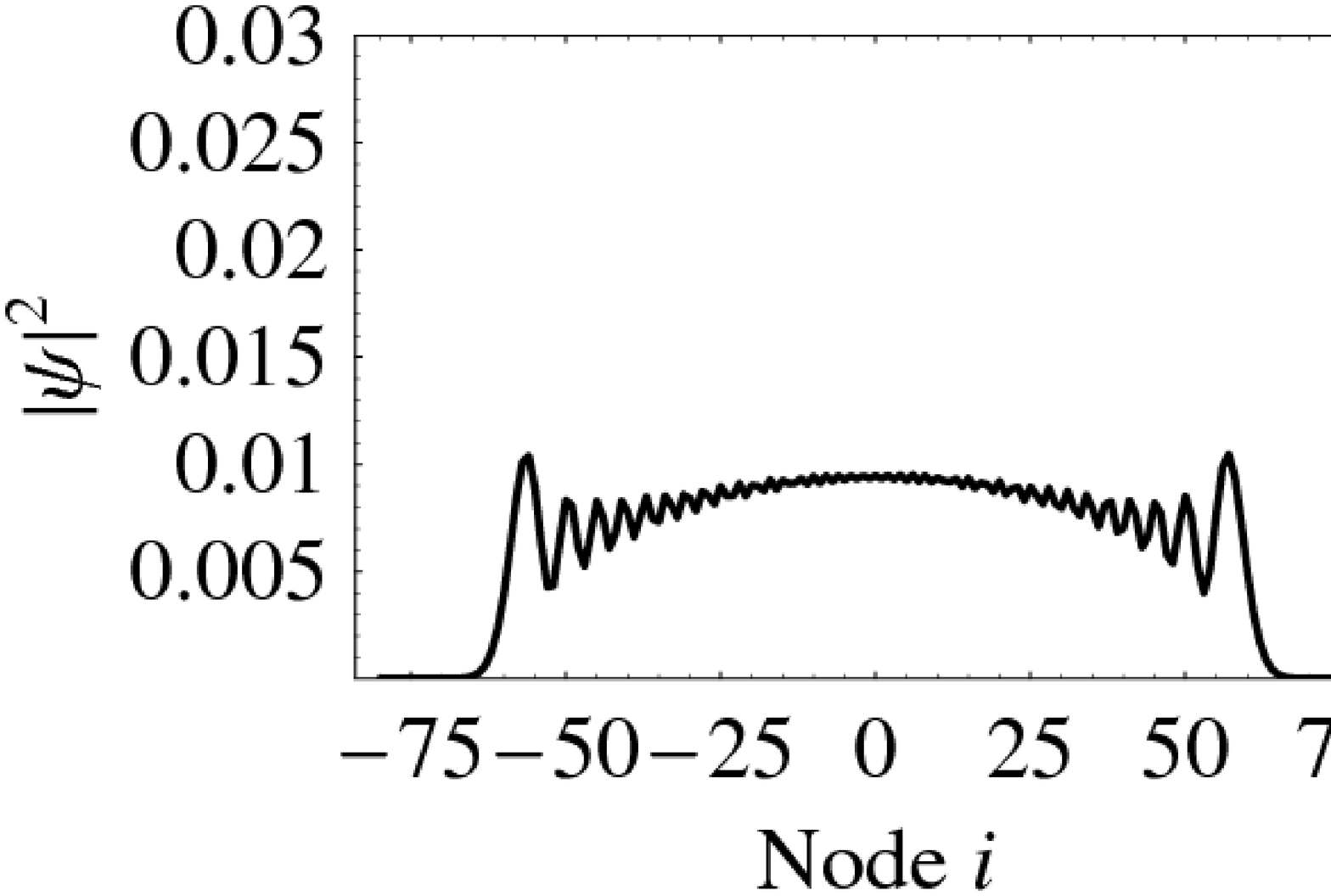}
    \includegraphics[width=5cm]{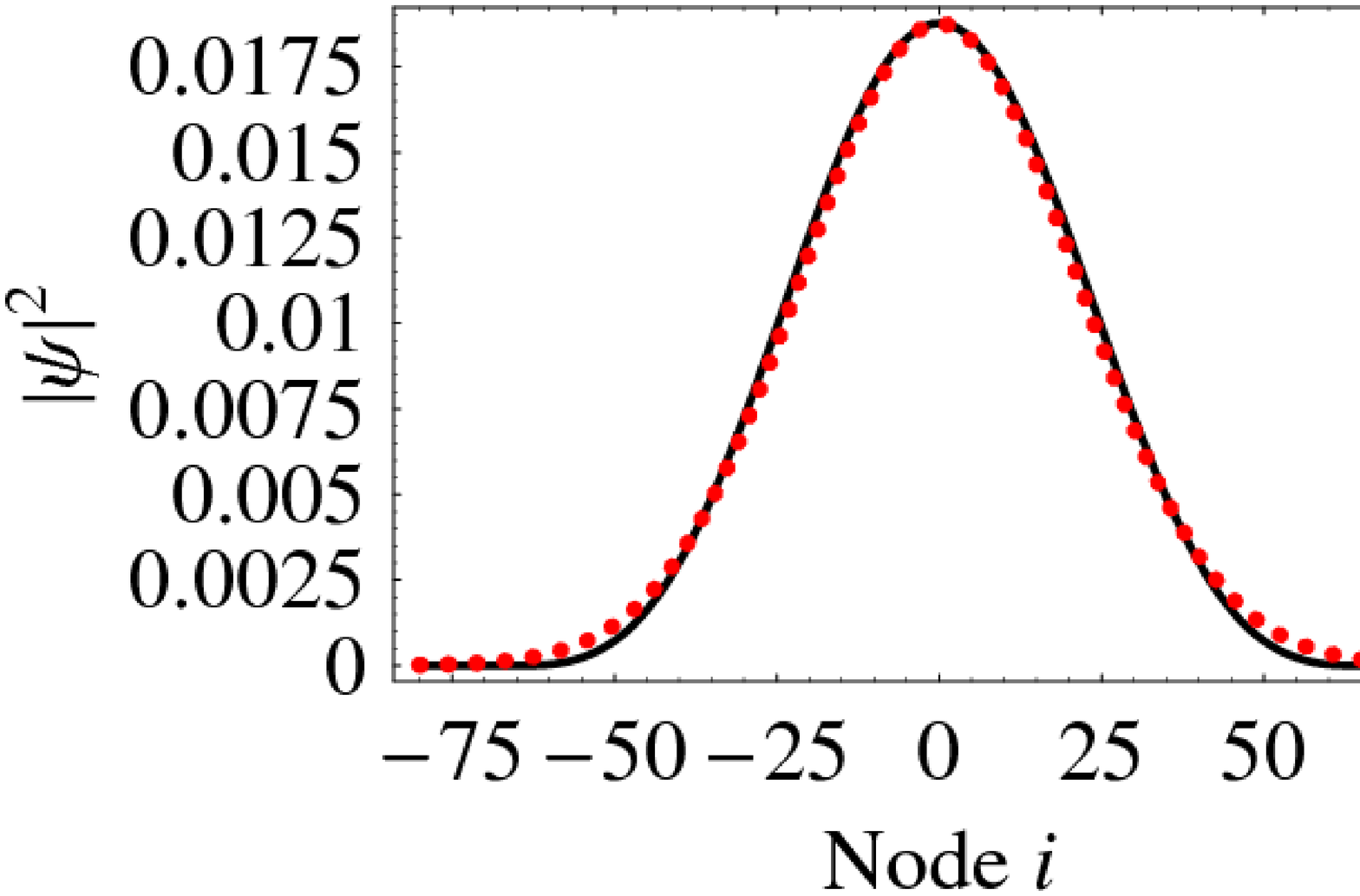}
    \caption{Convergence of the continuous-time quantum random walk probability
    distribution to a simple Gaussian for transition lengths $\lambda = 16, 4,
    3, 2,$ and $1$. The evolution is carried out using the coefficients of Eq.
    \ref{eqn.trans-rule-def-1} as the transition rates, with
    the initial condition $\psi(i=0,t=0)=1$ and evolution time $t=15$.}
    \label{fig.qrw-limit-1}
\end{figure}

\begin{figure}
    \centering
    \includegraphics[width=5cm]{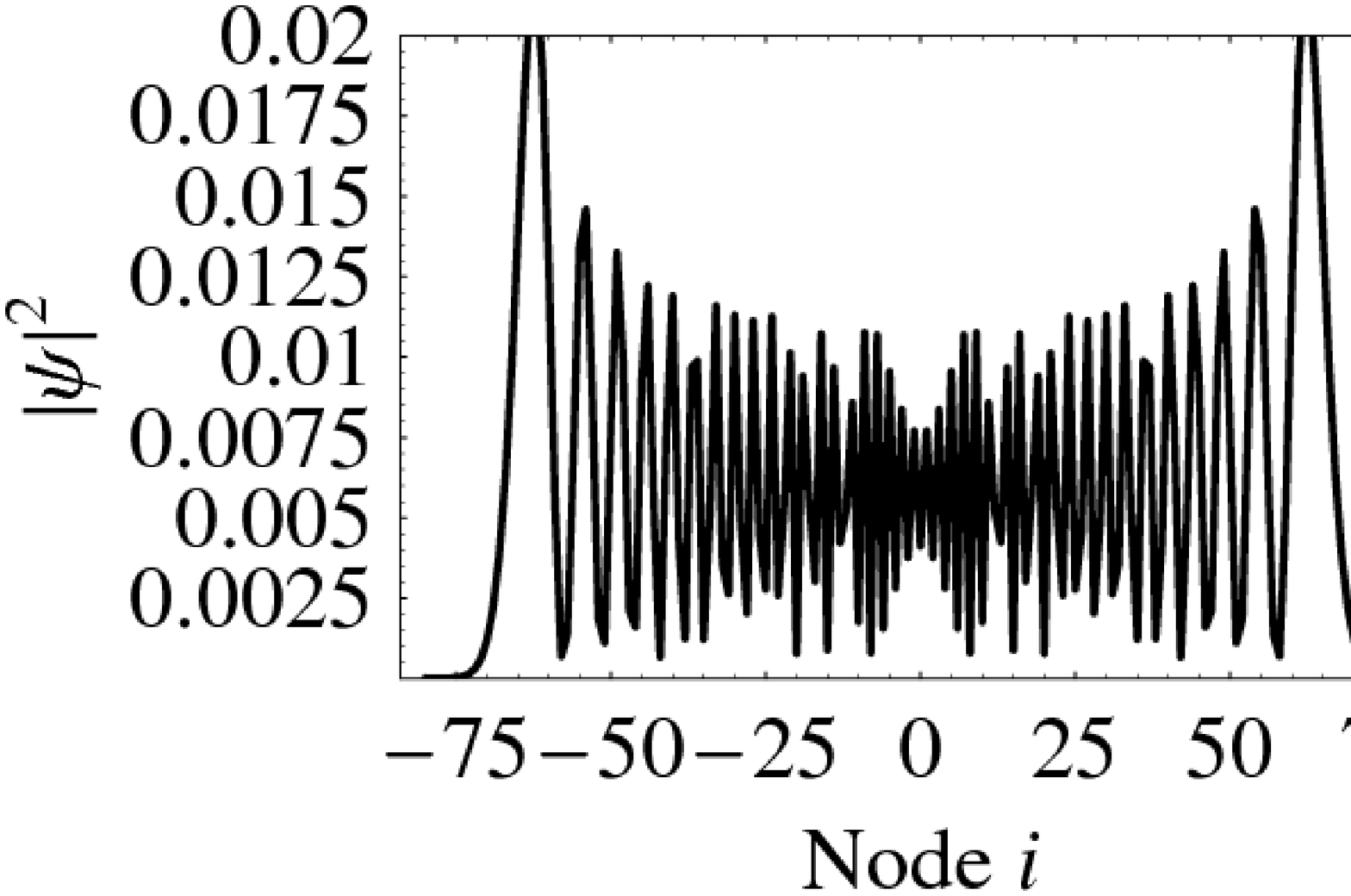}
    \includegraphics[width=5cm]{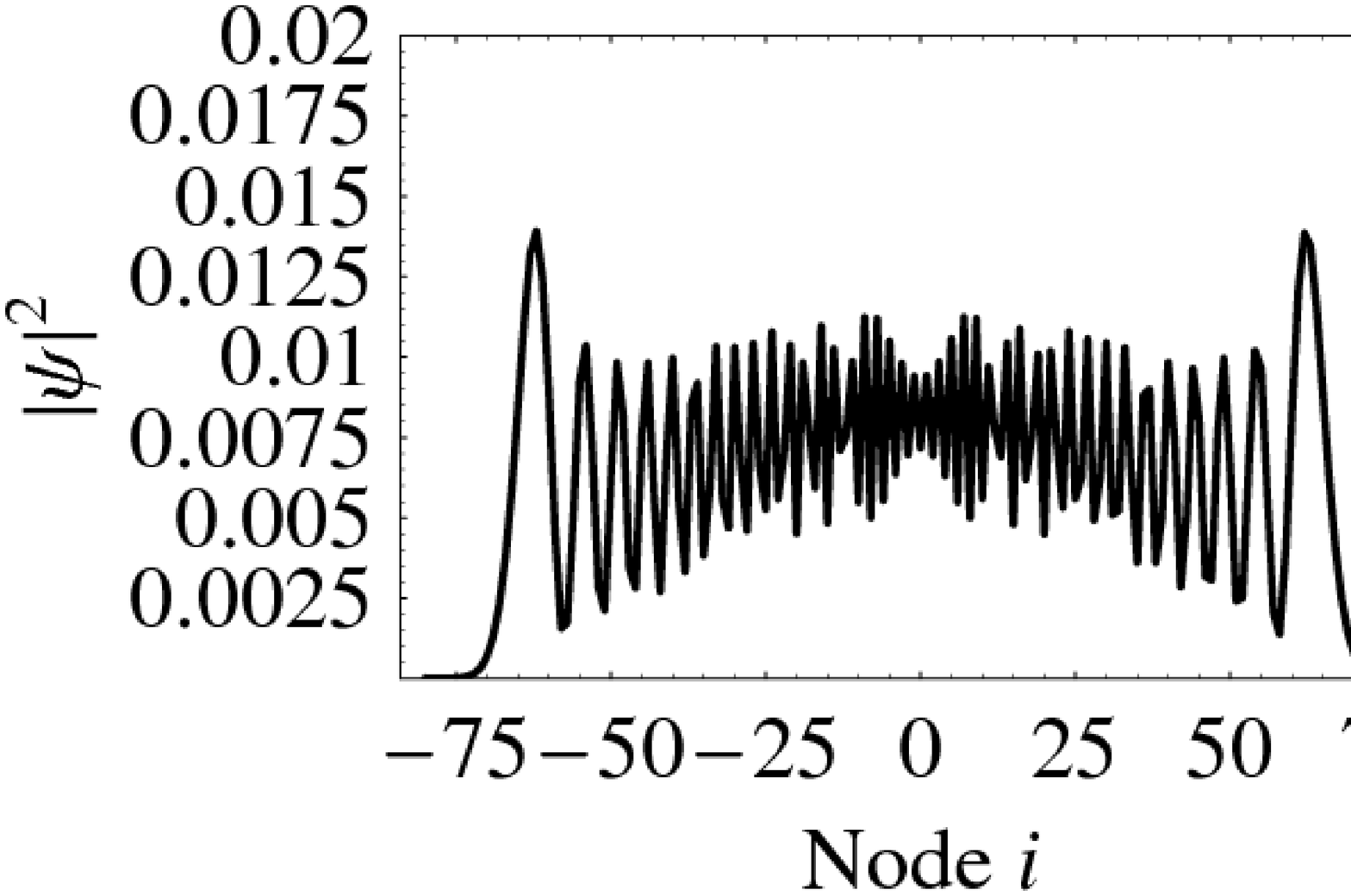}
    \includegraphics[width=5cm]{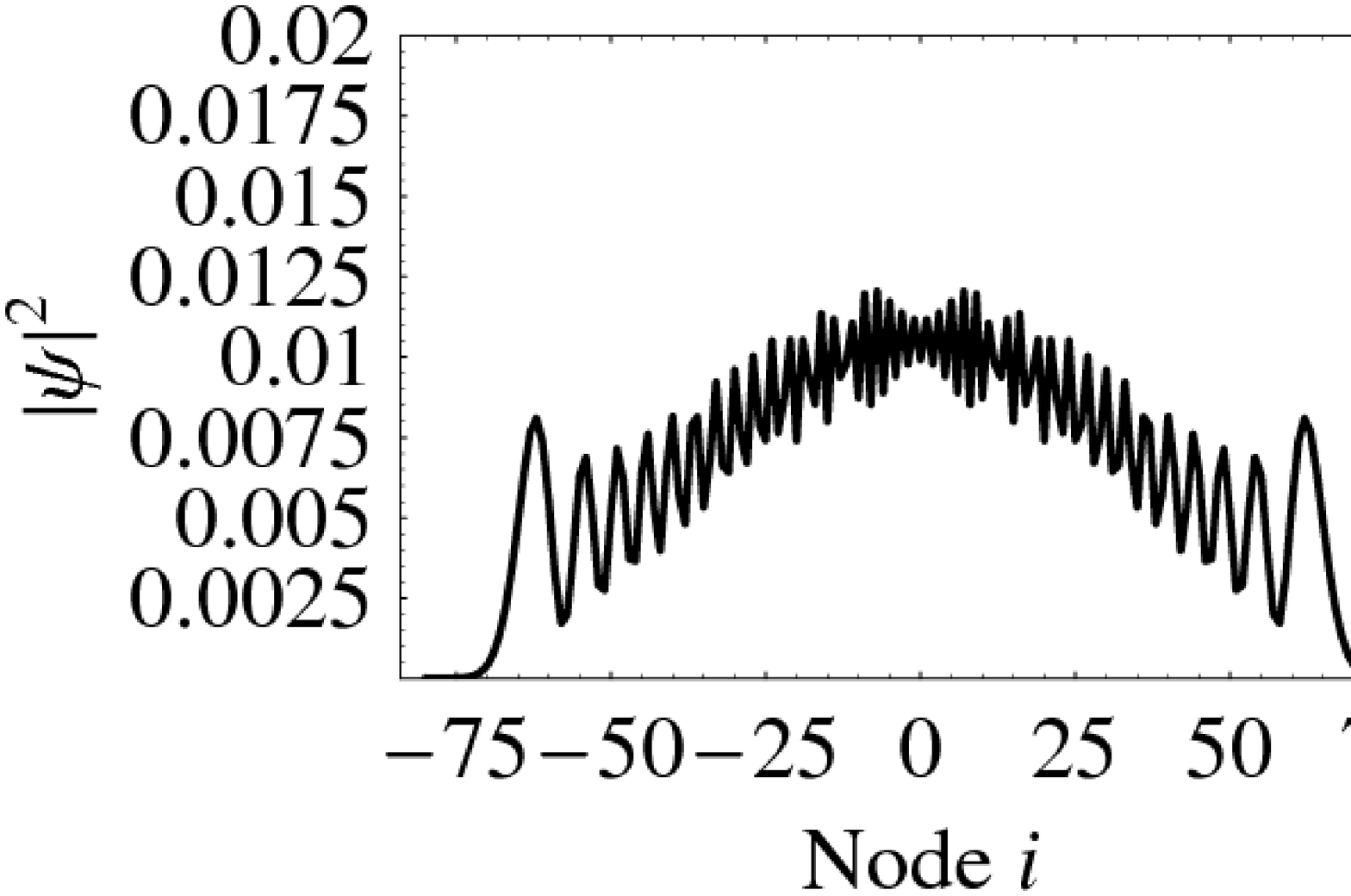}
    \includegraphics[width=5cm]{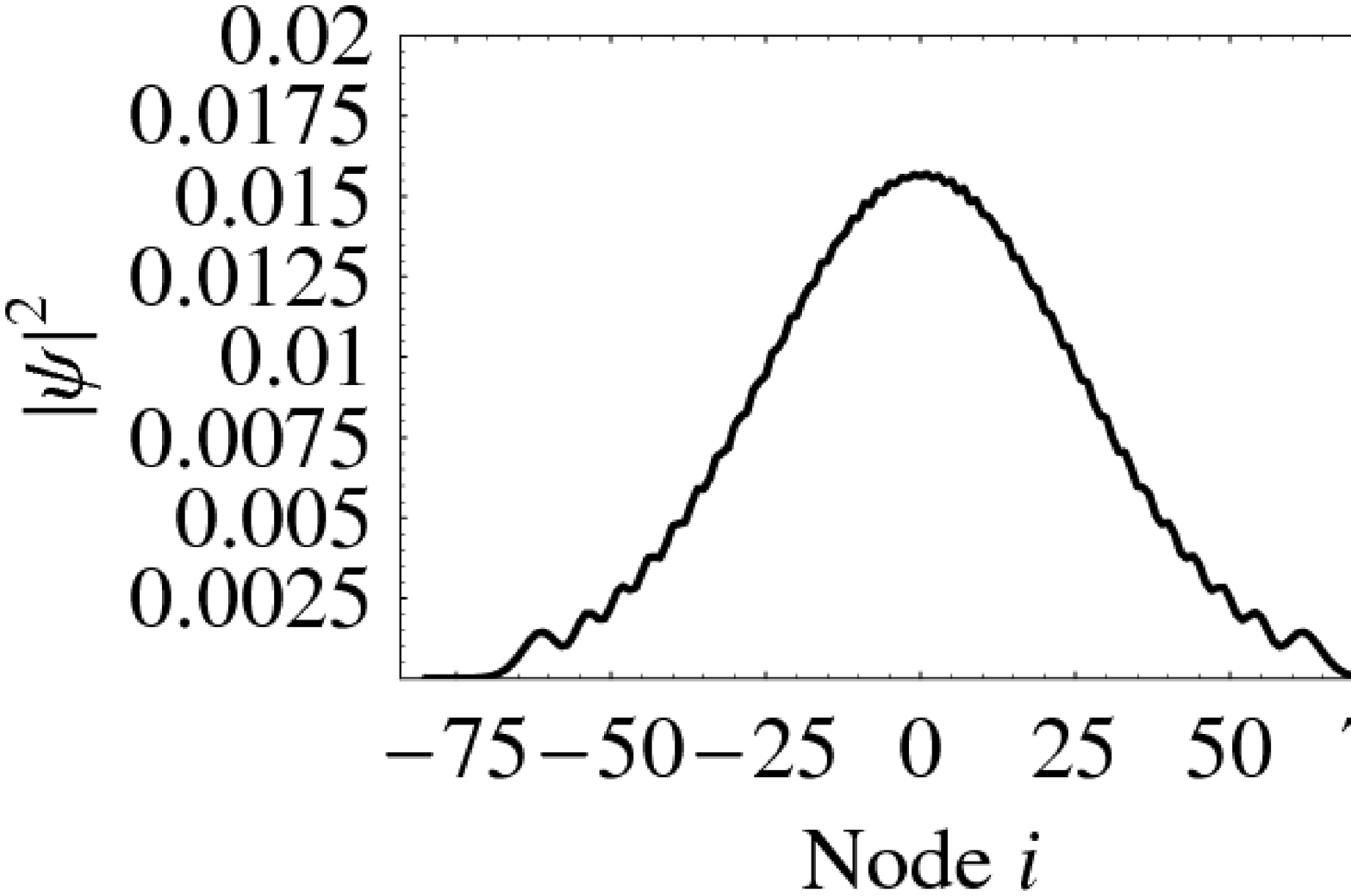}
    \includegraphics[width=5cm]{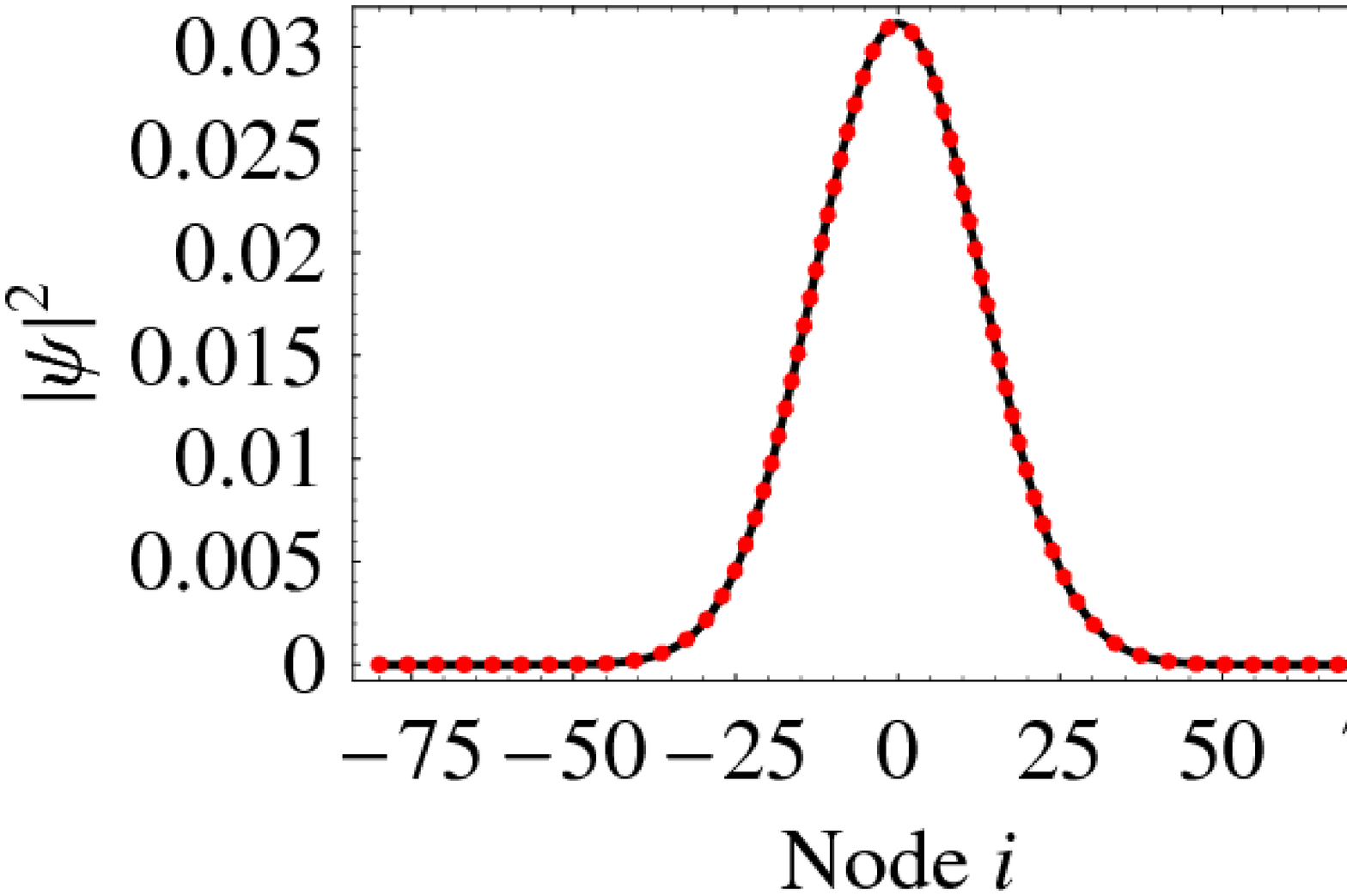}
    \caption{Convergence of the continuous-time quantum random walk probability
    distribution to a simple Gaussian for transition lengths $\lambda = 16, 4,
    3, 2,$ and $1$. The evolution is carried out using the coefficients of Eq.
    \ref{eqn.trans-rule-def-2} as the transition rates, with
    the initial condition $\psi(i=0,t=0)=1$ and evolution time $t=15$.}
    \label{fig.qrw-limit-2}
\end{figure}

Figures \ref{fig.qrw-limit-1} and \ref{fig.qrw-limit-2} show the
final probability distributions for $\psi(t=15)$ which is obtained
using Eq. \ref{eqn.distribution-amplitude-sum}. For the case
$\lambda=1$ we have also plotted the analytical solution for the
free propagation of a Gaussian wave packet \cite{book.Townsend1992}
\begin{equation}
    \psi_\mathrm{Gaussian}(x,t)=\frac{1}{\sqrt{\sqrt{2\pi} (\Delta x + \mathbbmtt{i}t/\Delta x)}} \exp\left(\frac{-x^2}{4(\Delta x^2+\mathbbmtt{i}t)}
    \right).
\end{equation}
Here we can see an almost perfect convergence of the quantum walk to
the familiar propagation of the Gaussian wave packet in free space.

\begin{figure}
    \centering
    \includegraphics[width=5cm]{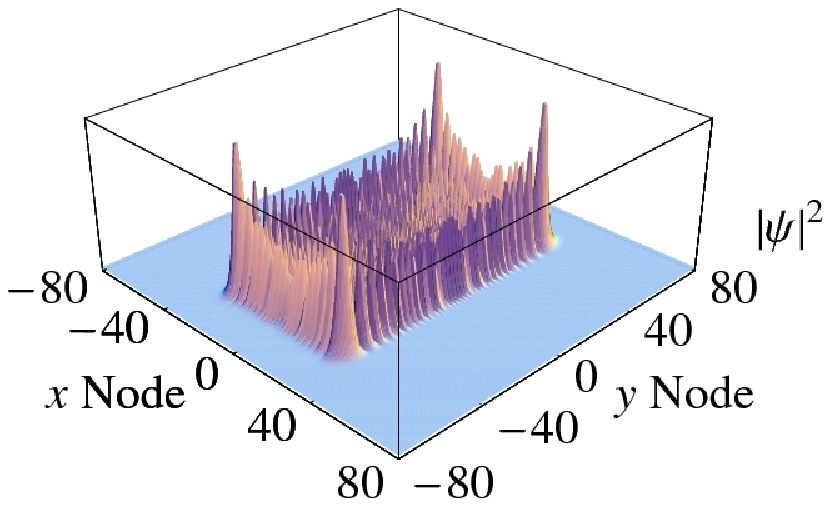}
    \includegraphics[width=5cm]{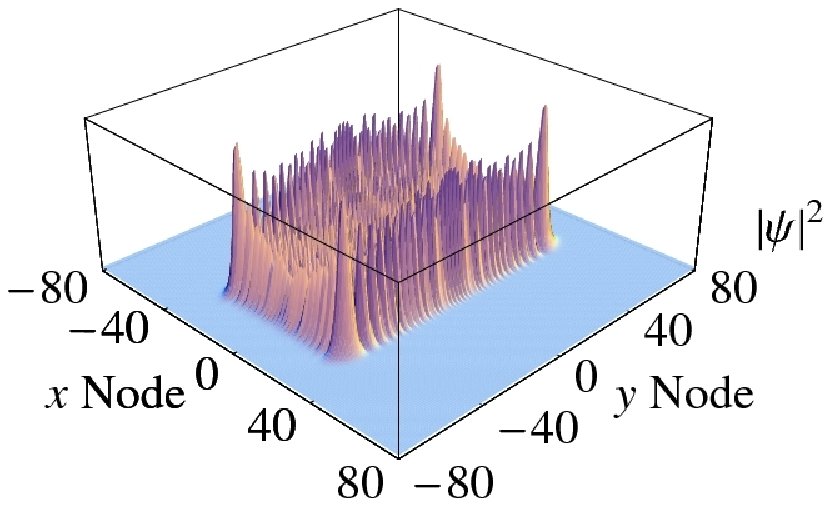}
    \includegraphics[width=5cm]{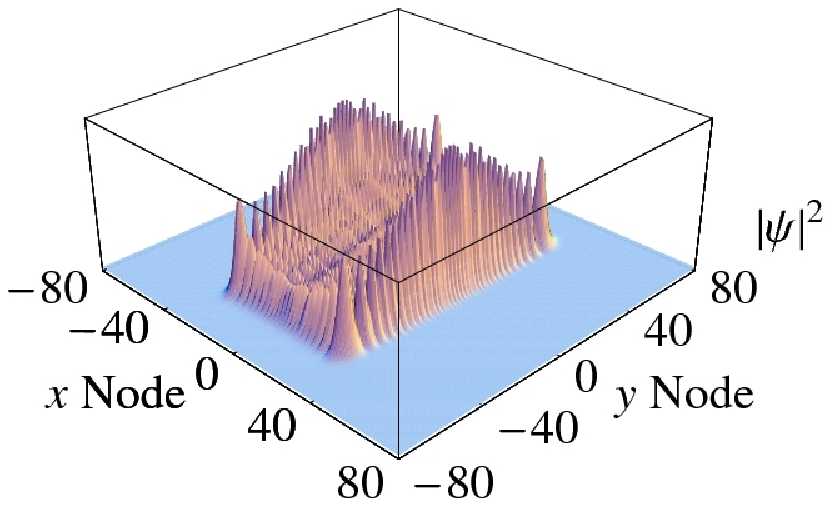}
    \includegraphics[width=5cm]{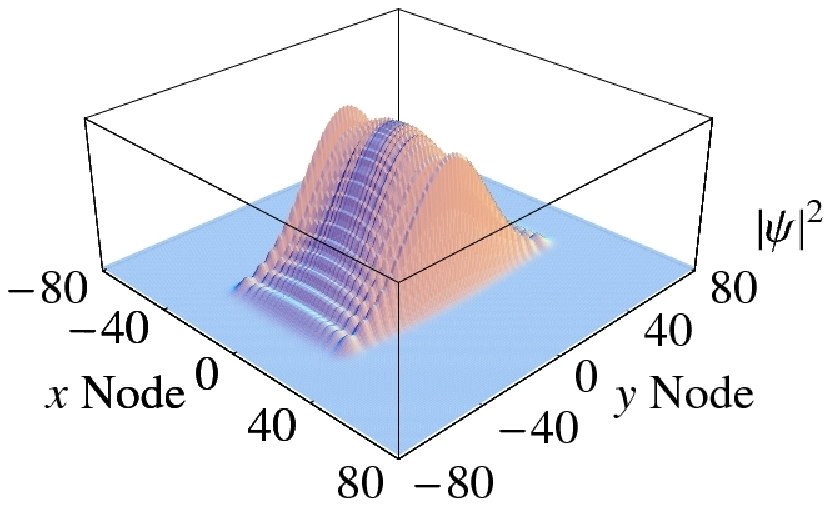}
    \includegraphics[width=5cm]{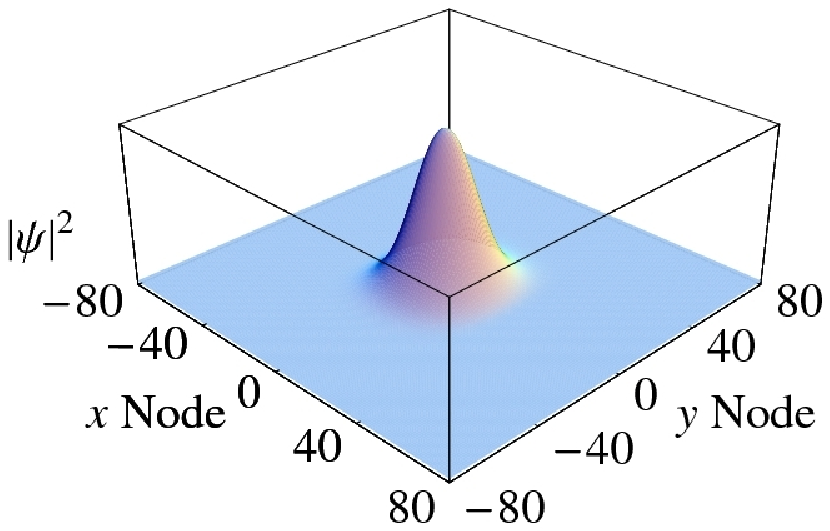}
    \caption{Convergence of a two-dimensional continuous-time quantum random walk probability
    distribution to a simple Gaussian for transition lengths $\lambda = 16, 4,
    3, 2,$ and $1$. The evolution is carried out using the coefficients
    of Eq. \ref{eqn.trans-rule-def-1} and \ref{eqn.trans-rule-def-2} as
    transition rates in the $x$ and $y$ directions respectively,
    with the initial condition $\psi_\text{2D}(i=0,j=0,t=0)=1$ and evolution time $t=15$.}
    \label{fig.qrw-2d-limit}
\end{figure}

Utilizing the computational scheme outlined in the
Appendix, extending the above analysis to higher dimensions becomes
trivial and computationally viable. Below is an example where we
have made the transition from discrete to continuous for a quantum
walk on a two-dimensional mesh. The transitions follow
$\mathcal{H}_1$ in the $x$-direction and $\mathcal{H}_2$ in the
$y$-direction. The resulting probability distribution for
$\psi_\text{2D}(t=15)$ is given in Figure
\ref{fig.qrw-2d-limit} where the quantum walk converges from its
characteristic evolution to a symmetric Gaussian distribution.

\section{Conclusion}
\label{section.conclusion}

We have shown that the evolutionary behavior of continuous-time
quantum random walks can be critically affected by the
discretization of state space. When the quantum walker has a finite
width, a virtually-discrete state space is constructed by keeping
the position states of the walk well separated so that their
corresponding distributions do not interfere. By allowing the
position states to move closer, approaching a continuum, the
walker's distribution at neighboring position states begin to
interfere which alters the evolutionary behavior. We have moreover
shown that when $\mathcal{H}\approx-\frac{1}{2}\nabla^2$ this can
even lead to the complete collapse of the quantum walk behavior to
that of a typical quantum wave propagation in free space.

Beyond theoretical interest however, our analysis find important
implications for the physical implementation of continuous-time
quantum random walks. Indeed a number of the existing experimental
schemes implicitly verify the notion that state space should be
discrete. Du et. al \cite{Du2003} demonstrated the implementation of
a quantum walk on a circle with four nodes, using a two-qubit NMR
quantum computer. Here the four dimensional state space spanned by
the spin states of the two qubits $\ket{\uparrow\uparrow},
\ket{\uparrow\downarrow}, \ket{\downarrow\uparrow}$ and
$\ket{\downarrow\downarrow}$, is clearly discrete in nature and the
walker has an infinitely narrow width.

Nevertheless the assumption that the walker has a finite width is in
fact a realistic one for many other physical systems such as a
single particle in real or momentum space, where a localized
distribution invariably arises from the fundamental uncertainty in
the particle's position for any given coordinates. In one such
proposal C\^{o}t\'{e} et al \cite{Cote2006} described a scheme based
on ultra cold Rydberg (highly excited) $^{87}$Rb atoms in an optical
lattice. In this scheme the walk is taking place in real space which
is clearly continuous, but the confinement of atomic wavefunction to
individual lattice sites amounts to a virtually-discrete state
space. In fact C\^{o}t\'{e} et al prescribe an even more
conservative condition: to eliminate the atoms except in every fifth
site (spacing 25$\mu$m) in order to achieve a better fractional
definition of the atom separation. Similarly Solenov and Fedichkin
\cite{Solenov2006} proposed using an a ring shaped array of
identical tunnel-coupled quantum dots to implement the
continuous-time quantum random walk on a circle. Given the
confinement of the electron wavefunction inside the individual
quantum dots, the authors have once again implicitly described a
virtually-discrete state space over the continuous real space.


\renewcommand{\theequation}{A-\arabic{equation}}
\setcounter{equation}{0}
\section*{APPENDIX: Computational Scheme}
\label{appendix.comp-scheme}

Here we present a computationally efficient scheme, referred to as
the \emph{Fourier-shift} method, which not only dramatically speeds
up the evaluation of matrix exponential in Eq. \ref{eqn.ext-qrw-cont-eqn}, but can also be
readily extended to higher dimensions when the connections between
the nodes naturally form a multidimensional mesh.

Our scheme is applicable to continuous-time quantum walks on $2d$
regular graphs. Specifically for a graph with $N$ vertices or nodes,
if we arrange all the nodes in a line indexed as $1,2,\ldots, N$, we
require that each node $i$ is connected to nodes $j = i-d,
i-d+1,\ldots, i+d-1, i+d$ with transition rates $\gamma_s$, where
$s=i-j$. Applying these transitions to the $i$th node we have
\begin{equation}
    \mathcal{H}\psi(i)=\sum_{s=-d}^d \gamma_{s}~\psi(i+s),
    \label{eqn.shrodinger-1}
\end{equation}
where we have made the time parameter $t$ implicit. Using the
notation
\begin{equation}
    \overset{\xrightarrow{~s~}}{(\psi)}_i \equiv \psi(i+s),
\end{equation}
where $\xrightarrow{~s~}$ represents shifting the entire vector
$\psi$ ($s$ nodes in the positive direction if $s>0$, $s$ nodes in
the negative direction if $s<0$, and no shift for $s=0$), we can
rewrite Eq. \ref{eqn.shrodinger-1} as
\begin{equation}
    \mathcal{H}\psi(i)=\sum_{s=-d}^d
    \gamma_{s}~\overset{\xrightarrow{~s~}}{(\psi)}_i =
    \left(\sum_{s=-d}^d
    \gamma_{s}~\overset{\xrightarrow{~s~}}{\psi}\right)_i,
    \label{eqn.shrodinger-2}
\end{equation}
which yields
\begin{equation}
    \mathcal{H} \psi=\sum_{s=-d}^d
    \gamma_{s}~\overset{\xrightarrow{~s~}}{\psi}.
    \label{eqn.shift-hamiltonian}
\end{equation}

This representation is desirable since vector shifts can be readily
expressed via the Fourier shift theorem, which in turn greatly
simplifies the evaluation of $\exp(-\mathbbmtt{i}\mathcal{H}t)$. The
discrete Fourier shift theorem states that
\begin{equation}
    \mathcal{F}_k\left\{\overset{\xrightarrow{~s~}}{\psi}\right\} =
    \mathcal{F}_k\left\{\psi\right\} p_k(s),
\end{equation}
where $\mathcal{F}_k $ represents the $k^\mathrm{th}$ element of the
discrete Fourier vector with $k=0, 1, 2 \ldots N-1$, and
\begin{equation}
    p_k(s)=\exp(2\pi \mathbbmtt{i} \widetilde{k} s/N),
    \label{eqn.fourier-elements}
\end{equation}
where $\widetilde{k}=k$ for $k \in [0, N/2]$, and
$\widetilde{k}=k-N$ for $k \in [N/2+1,N-1]$. Then applying the inverse transform we have
\begin{equation}
    \overset{\xrightarrow{~s~}}{\psi} = \mathcal{F}^{-1}\left\{\mathcal{F}\left\{\psi\right\} \otimes \mathcal{P}(s)
    \right\},
\end{equation}
where $\mathcal{P}(s)$ is a vector with elements $p_k(s)$ and
$\otimes$ is the direct vector product such that $(a_1, a_2, \ldots,
a_n)\otimes(b_1, b_2, \ldots, b_n)=(a_1b_1, a_2b_2, \ldots,
a_nb_n)$.

We can now use the above identity to rewrite Eq.
\ref{eqn.shift-hamiltonian} as
\begin{equation}
    \mathcal{H}\psi=
    \mathcal{F}^{-1}\left\{\mathcal{F}\left\{\psi\right\}\otimes
    \mathcal{Q}\right\},
\end{equation}
where
\begin{equation}
    \mathcal{Q}= \sum_{s=-d}^d \gamma_s ~ \mathcal{P}(s).
    \label{eqn.defn-Q}
\end{equation}
This results in the following simplification
\begin{align}
    \mathcal{H}^n~\psi & = \mathcal{F}^{-1}\left\{
    \ldots
    \mathcal{F}\left\{
    \mathcal{F}^{-1}\left\{\mathcal{F}\left\{\psi\right\}\otimes\mathcal{Q}\right\}
    \right\}\otimes\mathcal{Q}\ldots\right\} \\
    & =
    \mathcal{F}^{-1}\left\{\mathcal{F}\left\{\psi\right\}\otimes\mathcal{Q}^n\right\},
\end{align}
where $\mathcal{Q}^n$ denotes the direct product of $n$
$\mathcal{Q}$-vectors. Hence a polynomial expansion of Eq.
\ref{eqn.qrw-evolution} gives
\begin{align}
    \psi(t) & = \sum_{n=0}^\infty
    C_n (-\mathbbmtt{i}\mathcal{H}t)^n \psi(0) \\
    & = \mathcal{F}^{-1}\left\{\mathcal{F}\left\{\psi(0)\right\}\otimes
    \sum_{n=0}^\infty C_n (-\mathbbmtt{i}\mathcal{Q} t)^n \right\} \\
    & = \mathcal{F}^{-1}\left\{\mathcal{F}\left\{\psi(0)\right\}\otimes
    \mathcal{R}(t)
    \right\},
    \label{eqn.fourier-prop-method}
\end{align}
where $C_n$ are the expansion coefficients and elements $r_i(t)$ of
vector $\mathcal{R}(t)$ are related to elements $q_i$ of vector
$\mathcal{Q}$ via the relation $r_i(t)=\exp(-\mathbbmtt{i}q_i t)$.

Computationally this representation is much more efficient than a
direct evaluation of the $N \times N$ matrix
exponential in Eq. \ref{eqn.qrw-evolution}, since vector
$\mathcal{Q}$ needs to be calculated once, and for different values
of time $t$ we only require the evaluation of $N$ scaler
exponentials $\exp(-\mathbbmtt{i}q_i t)$ followed by the action of
$\mathcal{F}^{-1}\mathcal{F}$ which can be efficiently performed
using FFT.

\begin{figure}
    \centering
    \includegraphics[width=8cm]{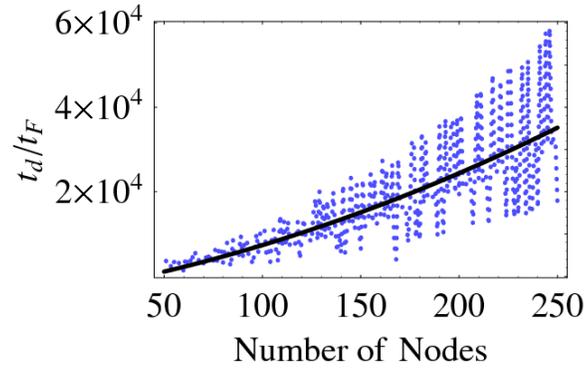}
    \caption{The computational efficiency of the Fourier-shift method over the direct method as a function of the number of nodes.
    The solid curve represents a quadratic fit to the efficiency data (dotted).}
    \label{fig.prop-efficiency}
\end{figure}

A numerical comparison between the direct and the Fourier-shift
method shows an excellent agreement in the resulting probability
distributions, while highlighting the Fourier method's tremendous
efficiency. Setting $N=200$ and using the transition
rates given by Eq. \ref{eqn.trans-rule-def-1}, we were able to
compute $\exp(-\mathbbmtt{i} \mathcal{H}_1 t)$ for $t=5$ and evolve
$\psi$ (initialized as a Gaussian distribution with $\Delta x = 2$)
with a relative accuracy better than $10^{-16}$. Repeating this for
$N$ ranging between 50 and 250, we also obtained an efficiency
factor $t_d/t_F$ as a function of $N$, where $t_d$ and $t_F$ are the
CPU times required for the direct and Fourier-shift methods,
respectively. This is plotted in Fig \ref{fig.prop-efficiency} along
with a quadratic fit to the data. The relatively large deviation of
the data from the mean is mainly due to the better optimization of
FFT packages for arrays of certain sizes, which should be considered
in the actual numerical implementation.

We can now extend Eq. \ref{eqn.fourier-prop-method} for application
to the quasi-continuous quantum walk on vector $\Psi$ where each
node is represented by a segment of $m$ elements in the vector. In
this case transition from node $i$ to node $j$ is equivalent to the
vector elements within the $i$th segment making a transition of
length $\lambda\abs{i - j}$. It is then easy to defining $(N \times
\lambda)$-element equivalents for vectors $\mathcal{P}(s)$,
$\mathcal{Q}$ and $\mathcal{R}$, represented by
$\mathcal{\overline{P}}(s)$, $\mathcal{\overline{Q}}$ and
$\mathcal{\overline{R}}$, and adjust the vector shifts by a factor
$\lambda$. More explicitly, the elements of
$\mathcal{\overline{P}}(s)$ are given by
\begin{equation}
    \overline{p}_k(s)=\exp(2\pi \mathbbmtt{i} \widetilde{k} s/\overline{N}),
\end{equation}
where $\overline{N}=N\lambda$, and
\begin{equation}
    \mathcal{\overline{Q}} = \sum_{s=-\lambda d}^{\lambda d} \gamma_s ~
    \mathcal{\overline{P}}(s).
\end{equation}
$\mathcal{\overline{R}}(t)$ is then constructed as before with
elements $\overline{r}_i(t)=\exp(-\mathbbmtt{i}\overline{q}_i t)$
and the time evolution of the quantum walk over $\Psi$ is simply
given by
\begin{equation}
    \Psi(t) = \mathcal{F}^{-1}\left\{\mathcal{F}\left\{\Psi(0)\right\}\otimes
    \mathcal{\overline{R}}(t)
    \right\}.
\end{equation}

\bibliography{cont-qrw}


\section{Figures}

\end{document}